\documentstyle[graphicx,pre,aps,float]{revtex}

\begin{document}
\draft

\title{Towards a realistic microscopic description of highway traffic}

\author{W~Knospe\dag, L~Santen\ddag, A~Schadschneider\S~and 
M~Schreckenberg\dag}

\address{\dag\ Theoretische Physik FB 10,
                  Gerhard-Mercator-Universit\"at Duisburg,
                  Lotharstr. 1, D-47048 Duisburg, Germany}
\address{\ddag\ Laboratoire de Physique Statistique,
                 \'{E}cole Normale Sup{\'{e}}rieure,
                 24, rue Lhomond, F-75231 Paris Cedex 05, France}
\address{\S\ Institut f\"ur Theoretische Physik,
                 Universit\"at zu K\"oln,
                 Z\"ulpicher Str. 77, D-50937 K\"oln, Germany}


\maketitle

\begin{abstract}
  Simple cellular automata models are able to reproduce the basic 
  properties of highway traffic. The comparison with empirical data for 
  microscopic quantities requires a more detailed description of the 
  elementary dynamics. 
  Based on existing cellular automata models we propose an improved discrete 
  model incorporating anticipation effects, reduced acceleration capabilities
  and an enhanced interaction horizon for braking. The modified model is 
  able to reproduce the three phases (free-flow, synchronized, and
  stop-and-go) observed in real traffic. Furthermore we find a good 
  agreement with detailed empirical single-vehicle data in all phases.
\end{abstract}


\vspace{1cm}

Cellular automata (CA) models for traffic flow \cite{nagel93}
allow for a multitude of new applications. Since their introduction it
is possible to simulate large realistic traffic networks using a 
microscopic model faster than real time \cite{esser,nagel99}. 
But also from a theoretical point of view, these kind 
of models, which belong to the class of one-dimensional driven lattice
gases \cite{sz}, are of particular interest. Driven lattice gases
allow us to study generic non-equilibrium phenomena, e.g.~boundary-induced 
phase transitions \cite{krug91}.
Now, almost ten years after the introduction of the first CA models, 
several theoretical studies and practical applications have improved the
understanding of empirical traffic phenomena (for reviews,
see e.g.~\cite{review,helbing_book,tgf97,tgf99,kerner_PW,kerner_TGF}).
CA models have proved to be a realistic description of
vehicular traffic, in particular in dense networks \cite{esser,nagel99}. 

Already the first CA model of Nagel and Schreckenberg~\cite{nagel93} 
(hereafter cited as NaSch model) leads to a quite realistic flow-density
relation (fundamental diagram). Furthermore, 
spontaneous jam formation has been observed. 
Thereby the NaSch model is a minimal model in the sense that any
further simplification of the model leads to an unrealistic behaviour. 
In the last few years extended CA models have been proposed which are 
able to reproduce even more subtle effects, e.g.~meta-stable states of
highway traffic~\cite{robert}. Unfortunately, the comparison of
simulation results with empirical data on a microscopic level is not 
that satisfactory. So far the existing models fail to reproduce 
the microscopic structure observed in measurements of real
traffic \cite{neubert}. But in highway
traffic in particular, a correct representation of the microscopic
details is necessary  
because they largely determine the stability of a traffic state and 
therefore also the collective behaviour of the system.

From our point of view a realistic traffic model should satisfy 
the following criteria. First it should reproduce, on a 
microscopic level, empirical data sets, and second, a very efficient
implementation of the model for large-scale computer simulations
should be possible. Efficient implementations are facilitated if a 
discrete model with local interactions is used. 

Our approach is based on a driving strategy which comprises four
aspects: 
\begin{itemize}
\item [(i)]  At large distances the cars move (apart from fluctuations)
with their desired velocity $v_{max}$. 
\item[(ii)] At intermediate distances drivers react to velocity changes of 
the next vehicle downstream, i.e.~to ``brake lights''. 
\item[(iii)] At small distances the drivers adjust their velocity such that 
safe driving is possible. 
\item[(iv)] The acceleration is delayed for standing
vehicles and directly after braking events.
\end{itemize}

These demands are incorporated by a set of update rules
for the $n$th car, characterized by its position $x_n(t)$
and velocity $v_n(t)$ at time $t$.
Cars are numbered in the driving direction, i.e.~vehicle $n+1$  
precedes vehicle $n$.
The gap between consecutive cars (where
$l$ is the length of the cars) is $d_n=x_{n+1}-x_n-l$, and $b_n$
is the status of the brake light (on (off) $\rightarrow$
$b_n=1 (0)$). 
In our approch the randomization parameter $p$ for the $n$th car can take on
three different values $p_0$, $p_d$ and $p_b$, depending on its current 
velocity $v_n(t)$ and the status $b_{n+1}$ of the brake light of the 
preceding vehicle $n+1$: 
\begin{equation}
p = p(v_{n}(t),b_{n+1}(t)) = \left\{\begin{array} {lll}
      p_b &{\rm if\ }b_{n+1}=1 {\rm \ and\ } t_{h} < t_{s}\\ 
      p_0 &{\rm if\ }v_n=0\\
      p_d &{\rm in\ all\ other\ cases}.
\end{array} \right.
\end{equation}
where $t_{h}$ and $t_{s}$ are defined later. The update rules then are
as follows: 

\begin{enumerate}
\item[0.] Determination of the randomization parameter:\\
$p = p(v_n(t),b_{n+1})$ 

\item[1.] Acceleration:\\
if $((b_{n+1} = 0)$ and $(b_{n} = 0))$ or
$(t_{h} \ge t_{s})$ then: \hspace*{0.2cm}$v_n(t+1) =
\min(v_n(t)+1,v_{max})$ 

\item[2.] NaSch braking rule:\\
$v_n(t+1) = \min(d_n^{(eff)},v_n(t))$ \\
if $(v_n(t+1) < v_{n}(t))$ then: \hspace*{0.2cm}
$b_n = 1$ 

\item[3.] Randomization, braking:\\
if $(rand() < p)$ then: \hspace*{0.2cm} $v_{n}(t+1)
=\max(v_n(t+1)-1,0)$\\ 
if $(p = p_{b})$ then: \hspace*{0.2cm} $b_{n} = 1$

\item[4.] Car motion:\\
$x_{n}(t+1) =x_{n}(t)+v_n(t+1)$\\
\end{enumerate}

The velocity of the vehicles is determined by steps $1-3$, while
step $0$ determines the dynamical parameter of the model. 
Finally, the position of the car is shifted in accordance to
the calculated velocity in step 4.

In the simulations we introduced a finer discretization than the one
usually used in the NaSch model. The length of the cells is given 
by $1.5$ m which leads to a velocity discretization of $5.4$ km h$^{-1}$.
Since one time step corresponds to $1$ s, 
this is slightly above a `comfortable' acceleration of about $\sim
1$ m s$^{-2}$~\cite{ite}. Each car has a length of $5$ cells.

Although the rules are formulated in analogy to existing models,
e.g. the NaSch model, some crucial differences exist. Compared to the
NaSch model with velocity-dependent randomization (VDR model)
\cite{robert}, where a velocity dependent randomization step has
already been
introduced, we even enhanced the action of the braking
parameter. Moreover our approach also includes anticipation effects
\cite{wolfgang,barret},
because the effective gap $d_n^{(eff)}$ is used instead of the real spatial 
distance to the leading vehicle. 

In order to illustrate the details of
our approach we now discuss the update rules stepwise.

\begin{enumerate}
\item[0.] The braking parameter $p$ is calculated. For a car 
at rest we apply the value $p=p_0$. Therefore $p_0$ determines the upstream
velocity of the downstream front of a jam. If the brake light of the 
car in front is switched on and it is found within
the interaction horizon we choose $p=p_b$. In all other cases we
choose $p = p_d$.

\item[1.] The velocity of the car is enhanced by one unit if the
car does not already move with maximum velocity. The car does not
accelerate if its own brake light or that of its predecessor
is on and the next car ahead is within the interaction horizon. 
  
\item[2.] The velocity of the car is adjusted according to the  
effective gap. 

\item[3.] The velocity of the car is reduced by one unit with a
  certain  
  probability $p = p(v_{n}(t),b_{n+1})$. If the car brakes due to the
  predecessor's brake light, its own brake light is switched on.

\item[4.] The position of the car is updated.

\end{enumerate} 

The parameters of the model are the following: the maximal velocity $v_{max}$,
the car length $l$, 
the braking parameters $p_d$, $p_b$, $p_0$, the cut-off time 
of interactions $h$ and the minimal security gap $gap_{security}$. 
The action of each parameter will be explained when comparing the
simulation results with the empirical data.

The two times $t_{h} = d_{n}/v_{n(t)}$ and  
$t_{s} = \min(v_{n}(t),h)$, where $h$
determines the range of interaction with the brake light, are introduced to
compare the time $t_{h}$ needed to reach the position of the leading vehicle
with a velocity-dependent (temporal) interaction horizon $t_{s}$.
$t_{s}$ introduces a cutoff which prevents drivers from reacting
to the brake light of a predecessor which is very far away.
Finally $d_n^{(eff)}=d_n+\max(v_{anti}-gap_{security}, 0)$ denotes the 
{\em effective} gap (where $v_{anti} = 
\min(d_{n+1},v_{n+1})$ is the expected velocity of the leading vehicle
in the next time step). The effectiveness of the anticipation is 
controlled by the parameter $gap_{security}$.
Accidents are avoided only if the constraint 
$gap_{security}\geq 1$ is fulfilled.

The parameter $h$ describes the horizon above which driving is not
influenced by the leading vehicle. Several empirical studies reveal
that $h$ corresponds to a {\em temporal} headway rather than to a
spatial one. The estimations for $h$ vary from $6$ s~\cite{george}, 
$8$ s~\cite{miller,schlums}, $9$ s~\cite{hcm} to $11$ s~\cite{edie}. 
%
Another estimation for $h$ can be obtained from the analysis of the 
perception sight distance. The perception sight distance is based on 
the first perception of an object in the visual field at which the driver 
perceives movement (angular velocity). 
In~\cite{pfefer} velocity-dependent perception sight distances
are presented which, for velocities up to $128$ km h$^{-1}$, larger than $9$s.
We therefore have chosen $h$ to be $6~s$ as a lower bound for the
time-headway.
Besides, our simulations show that correct results can only
be obtained for $h \ge 6$. This corresponds to a maximum horizon of $6
\times 20$ cells, or a distance of $180$ m, at velocity $v_{max}$.

\begin{figure}[h]
\begin{center}
	\includegraphics[width=11cm]{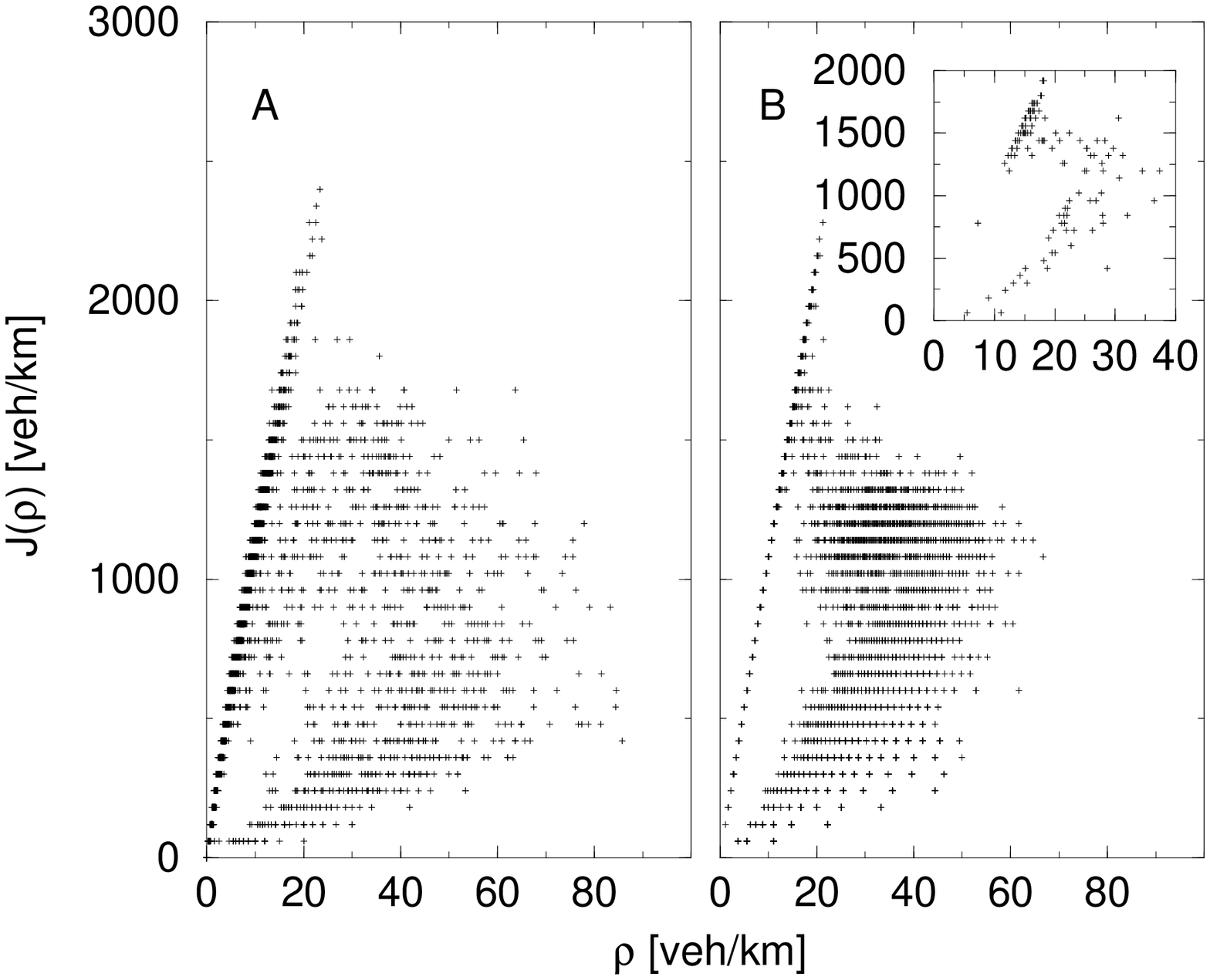}
	\includegraphics[width=11cm]{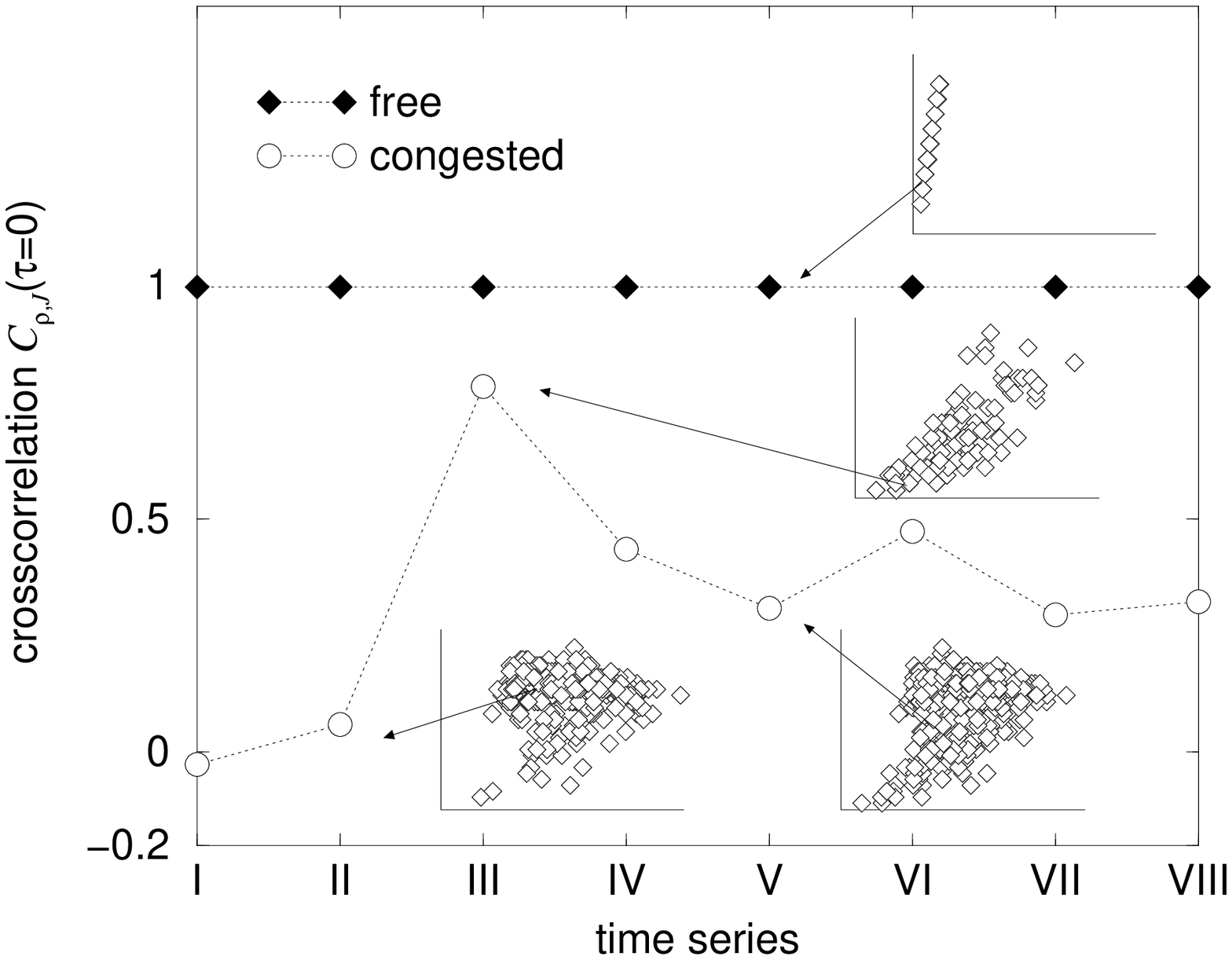}
\end{center}
\caption{Top: Comparison of the local fundamental diagram obtained by a
  simulation (B) with the corresponding empirical fundamental diagram (A)
  of [13]. 
  The inset shows the outflow of a megajam. 
  Bottom:
  Cross-covariance of the flow and the density for  
  different 
  system initialisations. The insets show the corresponding
  fundamental diagrams. The simulations
  are performed on a ring with a length of $50 000$ cells, each corresponding
  to $1.5$ m in reality. The parameters of
  the model are given by $v_{max} = 108$ km h$^{-1}$ $= 20$ cells s$^{-1}$,
$p = 0.1$, $p_0 = 0.5$, $p_b = 0.94$, $gap_{security} = 7$ and $h=
6$.} 
\label{carnaschlocal}
\label{crosscovariance}
\end{figure}

Analogously to the empirical setup the simulation data are evaluated
by an virtual induction loop, 
i.e. we measured the speed and the time-headway of the vehicles at a given
link of the lattice. This data set is analysed using the  
methods suggested in \cite{neubert}. 
In particular the density is calculated via the relation $\rho = J/v$ where
$J$ and $v$ are the mean flow and the mean velocity of cars passing the
detector in a time interval of $1$ min. 
This dynamical estimate of the density gives correct results only
if the velocity of the cars between two measurements is constant, but
for accelerating or braking cars, e.g.~in stop-and-go traffic, the
results do not coincide with the real occupation \cite{neubert}.  
In addition to the aggregated data, the single-vehicle data of each 
car passing the detector are also analysed.
Compared to the real part of the highway, analysed in \cite{neubert}, 
we performed our simulations on a simplified lattice, i.e.~we used 
a single-lane road with periodic boundary conditions. This is
justified because the empirical data do not show a significant lane
dependence.

\begin{figure}[hbt]
\begin{center}
	\includegraphics[width=11cm]{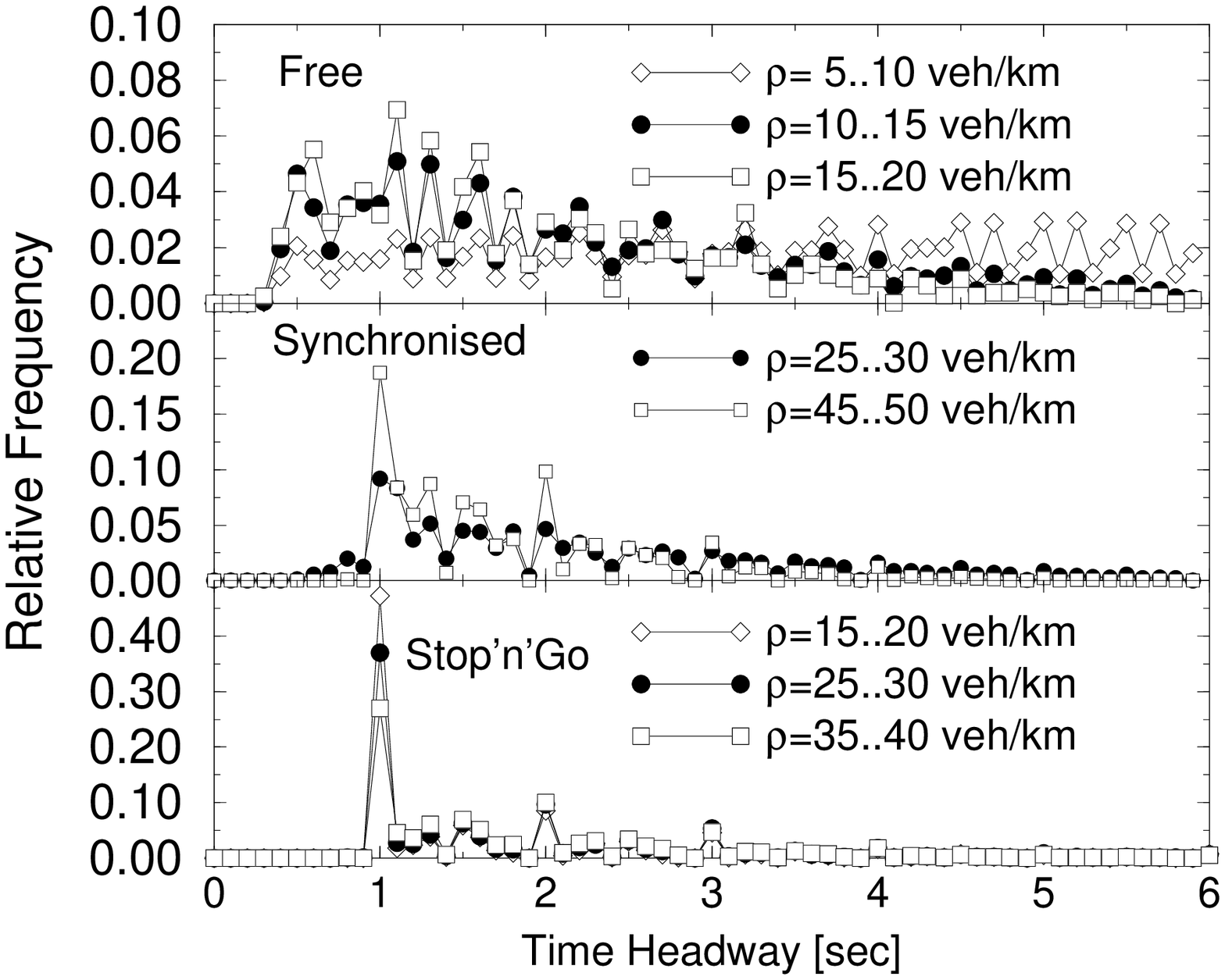}
	\includegraphics[width=11cm]{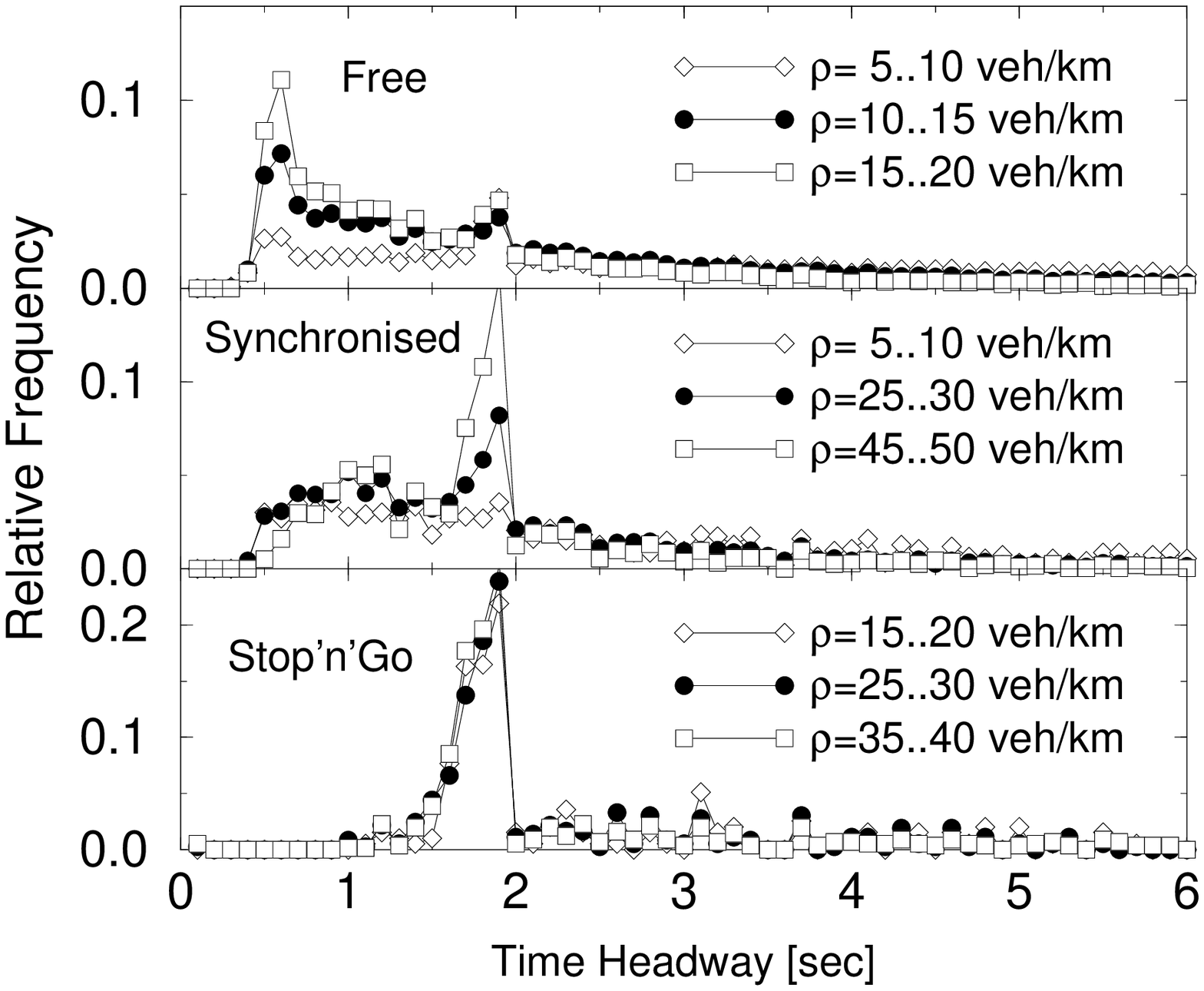}
\end{center}
\caption{Time-headway distribution for different density regimes 
        obtained from computer simulations (top) and from empirical
        data (bottom). The empirical data are reproduced from [13].
        For the simulations we used a periodic lattice and the same 
        set of parameters as in figure~\ref{carnaschlocal}.}
\label{carnaschmixed}
\end{figure}

The first goal of a traffic model is to reproduce in detail empirical
fundamental diagrams. This is obviously the case for 
the model proposed here (see figure \ref{carnaschlocal}).
The empirical fundamental diagram already allows us to
determine some of the parameters of the model. The slope of the 
fundamental diagram in the free-flow regime determines $v_{max}$. 
Moreover the maximum value of the flow is largely determined 
by the braking parameter $p$.
For the finer discretization used in the simulations
we found a reasonable agreement for $v_{max} = 20$. 
The simulated densities are less distributed than in 
the empirical data set. This observation 
is simply an artifact  
of the discretization of the velocities which
determines the upper limit of detectable densities.

The next parameter which can be directly related to an empirical
observable quantity is the braking parameter $p_0$. This parameter
determines the outflow of a jam and therefore its upstream 
velocity.  We used the value $p_0 = 0.5$ which leads
to an upstream velocity of a compact jam of approximately $12.75$ km h$^{-1}$.
This velocity is also in accordance with empirical
results~\cite{kerner96}. Moreover 
empirical work has revealed that the outflow of a jam is
smaller than the maximal possible flow under free-flow conditions
\cite{kerner96}.
The inset of figure \ref{carnaschlocal} shows that this is well
reproduced by the simulations.

A more detailed statistical analysis \cite{neubert} of the time-series 
of flux, velocity and density allows for the identification of three 
different traffic states \cite{kerner96,kerner_prl81,kerner_prl79}.
Following the arguments of \cite{neubert} free-flow and congested 
traffic states can be identified by means of the average velocity. 
A contiguous time-series of minute averages above $90$ km h$^{-1}$ 
was classified as free flow, otherwise as congested flow.
In figure \ref{crosscovariance} the cross-covariance 
\begin{equation}
cc(J,\rho)=\frac{\langle J(t)\rho(t+\tau) \rangle -\langle
 J(t)\rangle\langle \rho(t+\tau)\rangle} {\sqrt{\langle J^2(t) \rangle
 -\langle J(t)\rangle^2}\sqrt{\langle \rho^2(t) \rangle -\langle
 \rho(t)\rangle^2}}
\end{equation}
of the flow and the local measured density for different traffic states
is also shown. 
In the free-flow regime the flow is strongly coupled to the density 
indicating that the average velocity is nearly constant.
Also for large densities, in the stop-and-go regime, the flow is mainly
controlled by density fluctuations. In the mean density region there is a 
transition between these two regimes.
At cross-covariances in the vicinity of zero the fundamental diagram shows a
plateau. Traffic states with $cc(J,\rho) \approx 0$
were identified as synchronized flow 
\cite{neubert}. 
In the further comparison 
of our simulation with the corresponding empirical data we  
used these traffic states for synchronised flow data and congested states 
with $cc(J,\rho) > 0.7$ for stop-and-go data. 
The results show that the approach leads to realistic results for
the fundamental diagram and that the model is able to reproduce the 
three different traffic states.

\begin{figure}[!t]
\begin{center}
	\includegraphics[width=11cm]{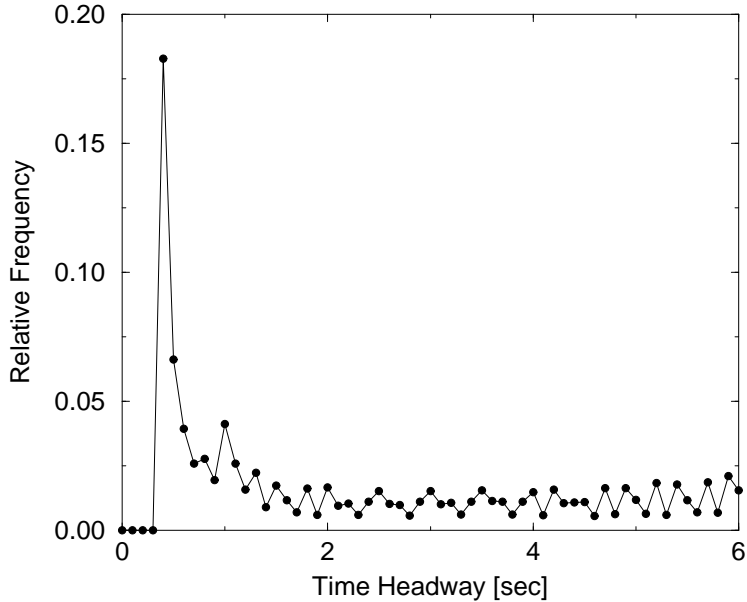}
\end{center}
\caption{Time-headway distribution in the free flow regime of a 
        system with 
	different types of
        vehicles. The maximal velocity of the slow vehicles is 
        $v_{max} = 108$ km h$^{-1}$ $= 20$ cells s$^{-1}$, and  of
the fast vehicles  
        is $v_{max} = 135$ km h$^{-1}$ $= 25$ cells s$^{-1}$.  We
considered $25 \% $ 
        of the vehicles as 
  fast vehicles (note that these are  vehicles
  which disregard the speed limit). The simulations were performed on 
  a lattice of size $L = 100 000$. Apart from that, the same set 
  of parameters as in figure~\ref{carnaschlocal} was used.}
\label{carnaschtime}
\end{figure}

Next we compare the empirical data and simulation results on a
microscopic level. Measurements of the time-headway distributions 
are the microscopic equivalents to measurements of the flow.
In figure \ref{carnaschmixed} the simulated and empirical 
time-headway distributions for different 
density regimes are shown. The time-headways 
obtained from Monte Carlo simulations
are calculated via the relation
$ \Delta t = {\Delta x}/{v} $ with a resolution of $0.1$ s.
Due to the discrete nature of the model, large fluctuations occur. In the 
{\it free-flow state} extremely small time-headways have been found,
in accordance with the empirical results. Nevertheless, for our
standard simulation setup the statistical weight of these small
time-headways is significantly underestimated. Therefore we 
also performed simulations of 
a system with different types of
vehicles. Figure \ref{carnaschtime} shows that the
origin of the peak at small time-headways are fast cars driving in 
small platoons behind slow vehicles. The location of this peak is mainly
influenced by the parameter $gap_{security}$. The results for congested 
flow, however, are not influenced by different types of vehicles.

\begin{figure}[!t]
\begin{center}
	\includegraphics[width=11cm]{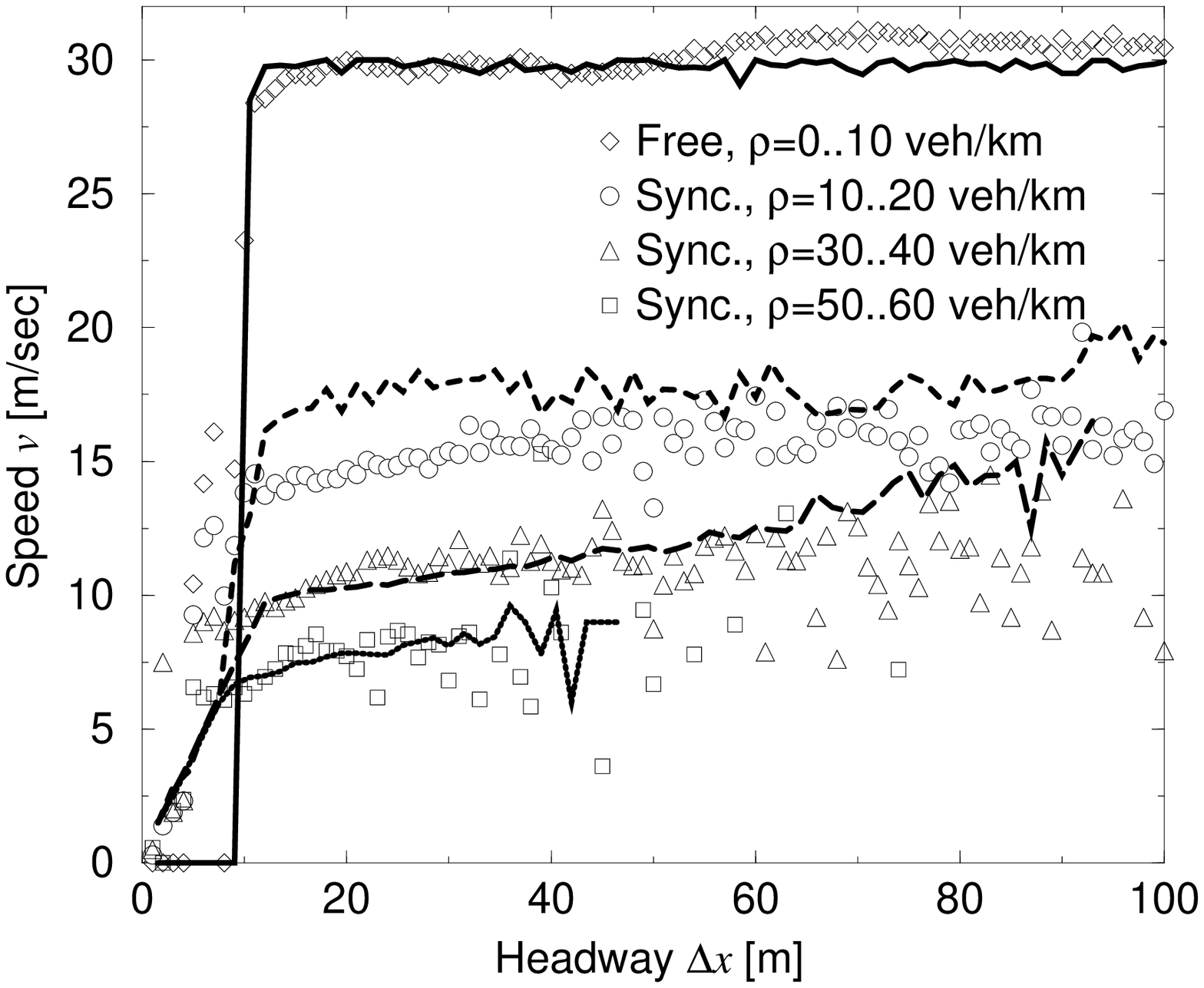}
	\includegraphics[width=11cm]{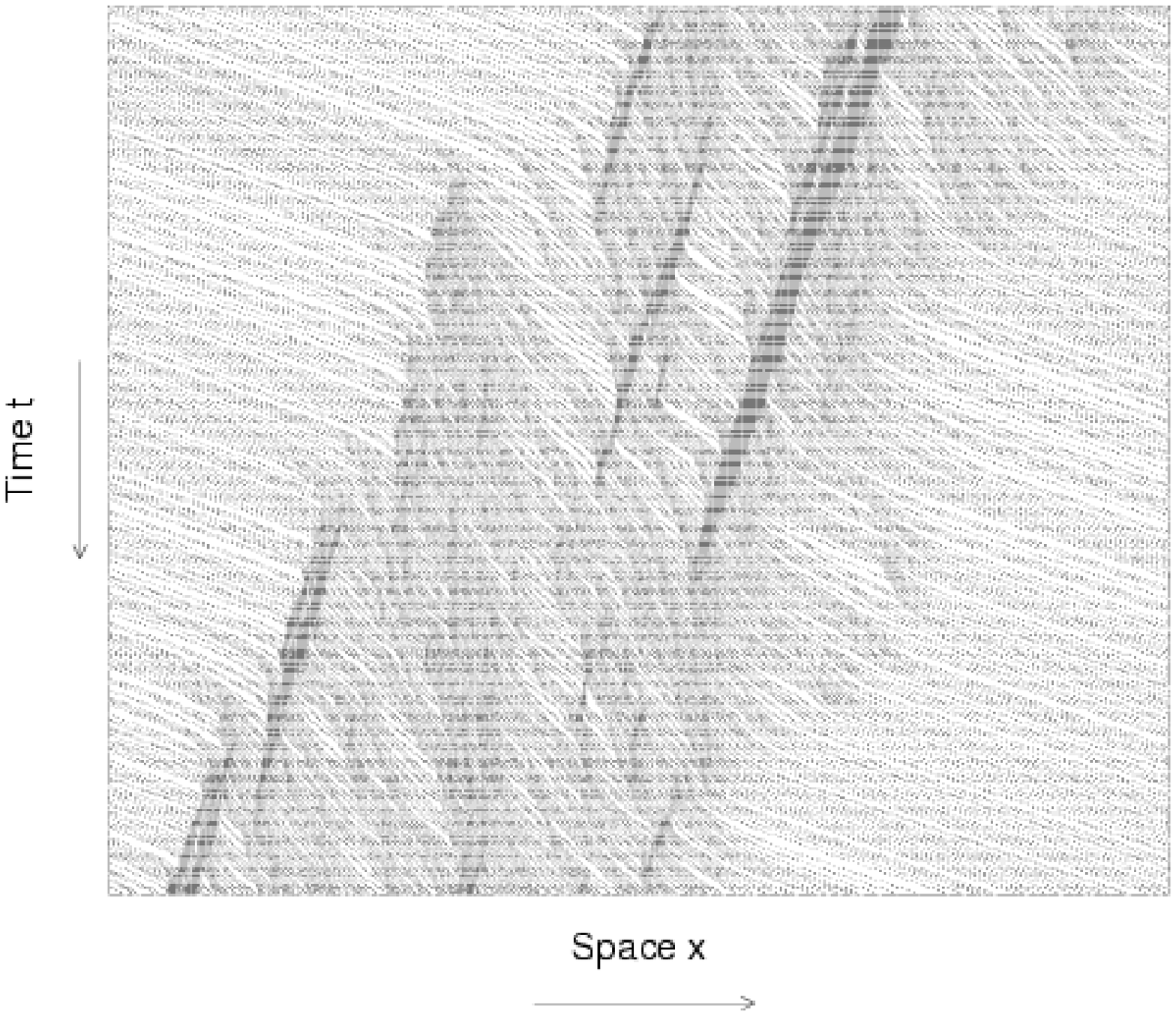}
\end{center}
\caption{Top: The mean speed chosen by the driver as a function of the gap
  to his predecessor. Comparison of 
  simulations (lines) with empirical data (symbols) [13].
  The system size and the parameters comply with figure \ref{carnaschlocal}. 
 Bottom: Space-time plot of a periodic system for a global density of
  $27$ vehicles km$^{-1}$. The cars are moving from left to
right. Note that the cars have 
        a length of $5$ cells.}
\label{carnaschov}
\label{spacetime}
\end{figure}

The ability to anticipate the predecessor's 
behaviour becomes weaker with increasing density so that small 
time-headways almost vanish in the synchronized and stop-and-go state.
Two peaks arise in these states: The peak at a time of $2$ s can be
identified with the driver's efforts for 
safety: it is recommended to drive with a distance of $2$ s. 
Nevertheless, with increasing density the NaSch peak at a time of $1$ s (in
the NaSch model the minimal time-headway is restricted to $1$ s) becomes
dominant. This result is also due to the discretisation of the model
which triggers the spatial and temporal distance between the cars.
Therefore the continuous part of the empirical distribution shows 
a peaked structure for the simulations.

For the correct description of the car-car interaction the distance-headway
(OV-curve) gives the most important information for the adjustment of 
the velocities~\cite{bando}. 
For densities in the free-flow regime it is obvious that the OV-curve
(figure \ref{carnaschov})
deviates from the linear velocity-headway curve of the NaSch model.
Due to anticipation effects, smaller distances occur so that driving with
$v_{max}$ is possible even within very small headways.
This strong anticipation becomes weaker with increasing density 
and cars tend to have smaller velocities than the 
headway allows so that the OV-curve saturates for large distances.
The saturation of the velocity, which is characteristic for synchronized 
traffic, was not observed in earlier approaches. The value of the 
asymptotic velocities can be adjusted by the last free parameter $p_b$.

The simulation results of the approach show that the empirical
data are reproduced in great detail. We observed three qualitatively
different microscopic traffic states which are in accordance with 
the empirical results (see the space-time plot in
figure \ref{spacetime}). The deviations of the simulation results  
are mainly due to simple discretization artifacts which do not reduce
the reliability of the simulation results. 
We also want to stress the fact 
that the agreement is on a microscopic level.

This improved realism of our approach leads to a larger complexity of
the model compared to other models of this type~\cite{prep}. 
Nevertheless, due to 
the discreteness and the local car-car interactions, very
efficient implementations should still be possible. Moreover, the adjustable
parameter of the model can be directly related to empirical quantities. 
The detailed description of the microscopic dynamics will also lead
to a better agreement of simulations with respect to empirical data for
macroscopic 
quantities, e.g., jam-size distributions.
Therefore we believe that this approach should allow for 
more realistic micro-simulations of highway networks.\\

The authors have benefited from discussions with L.~Neubert,
R.~Barlovi\'{c} and T.~Huisinga. L.~Santen~acknowledges support
from the Deutsche Forschungsgemeinschaft under Grant No.~SA864/1-1.
We also thank the Ministry of Economic Affairs, Technology
and Transport of North-Rhine Westfalia as well as to the Federal Ministry of
Education and Research of Germany for  financial support (the latter within
the BMBF project ``SANDY'').

\end{document}